\documentclass[12pt]{article}
\usepackage[dvips]{color}
\usepackage{epsfig}
\usepackage{amsmath}
\usepackage{graphicx}

\begin{document}
\title{Corrections Terms for the Thermodynamics of a Black Saturn}
\author{{Mir Faizal$^{a}$\thanks{Email:
f2mir@uwaterloo.ca} and Behnam Pourhassan$^{b}$\thanks{Email:
b.pourhassan@du.ac.ir}}\\
$^{a}${\small {\em Department of Physics and Astronomy, University of Waterloo, Waterloo, ON N2L 3G1, Canada}}\\
$^{b}${\small {\em  School of Physics, Damghan University, Damghan, Iran}}}
\date{}
\maketitle
\begin{abstract}
In this paper, we will analyze  the effects of thermal  fluctuations on the stability of a black Saturn.
The entropy of the black Saturn will get corrected due to these thermal fluctuations. We will demonstrate
that the correction term generated by these thermal fluctuations is a logarithmic term.
Then we  will use this corrected value of the entropy to obtain   bounds for various
parameters of the black Saturn.
We will also analyze the thermodynamical stability of the black Saturn in presence of thermal fluctuations,
using this corrected value of the entropy.
\end{abstract}
\tableofcontents
\section{Introduction}
If entropy is not associated with a black hole, then the entropy of the universe will spontaneous reduce whenever an object with a finite entropy crosses the horizon. Thus, entropy has to be associated with a black hole to prevent the violation of the second law of thermodynamics \cite{1, 1a}. In fact,
black holes have more entropy than any other object of  the same volume \cite{2, 4}.
This prevents the violation of second law of thermodynamics. This maximum  entropy of the black holes is  proportional to the area  of the horizon \cite{4a}. Thus, if $S$ is the entropy associated with a black hole, and $A$ is the area of the horizon, then the relation between $S$ and $A$ can be expressed as  $S = A/4$. The observation that the entropy scales with the area of the black hole, instead of its volume, has motivated the development of the holographic principle~\cite{5, 5a}.
The holographic principle  states that the degrees of freedom in a region of space is the same as the degrees of freedom on the boundary surrounding that region of space.
The geometry of black holes will undergo quantum fluctuations. These quantum corrections will lead to thermal fluctuations. These thermal fluctuations will in turn generate correction terms for various thermodynamical quantities associated with black holes \cite{l1, SPR}.
Thus the holographic principle can get modified near Planck scale \cite{6, 6a}. It may be noted that even though
the thermodynamics of black holes is expected to get corrected due to thermal fluctuations, we can
neglect such correction terms for large black holes. This is because these thermal fluctuations occur because
of quantum fluctuations of the geometry of space-time, and such quantum fluctuations can be neglected for large black
holes. However, as the black holes radiate Hawking radiation, they
tend to evaporate in course of time. They the size of the black holes reduces in course of time due to
the Hawking radiation. As the black holes become smaller the quantum fluctuation give more dominate contributions
to the geometry of space-time. Thus, the thermal fluctuations cannot be neglected for small black holes, or for
black hole at the last stages of their evaporation. The  correction terms to the entropy of black holes coming from
thermal fluctuations have been calculated. It has been demonstrated that these correction terms are expressed as logarithmic functions of the original thermodynamic quantities.

The corrections to the thermodynamics of black holes has also been calculated using
the density of microstates for asymptotically flat black holes \cite{1z}. This analysis has been done
in the framework of non-perturbative quantum  general
relativity. Here  conformal blocks of a well defined  conformal field theory
is associated with the density of states for a black hole.
This density of states is then used to calculate the relation between the entropy of a black hole and
the area of its horizon. The leading order relation between the  entropy of a black hole and
the area of its horizon is observed to be the standard Bekenstein entropy area relation for the large black holes. However, this relation between the area and entropy of a black hole gets corrected in this analysis. The leading order   corrections  terms to the entropy of the black hole are demonstrated to  be   logarithmic corrections. It may be noted that the such corrections terms have also been calculated using the Cardy formula \cite{card}. In fact, it has been demonstrated using this formula that such logarithmic corrections terms will be generated for all black holes whose microscopic degrees of freedom are described by a conformal field theory. The correction terms to the entropy of a BTZ black hole have been calculated using such logarithmic exact
partition function \cite{gks}. It has been again observed that these correction terms can be expressed
using logarithmic functions. It has also been possible to obtain  logarithmic correction terms for the entropy
of a black hole by analyzing matter fields in backgrounds of a black hole  \cite{other, other0, other1}.

The correction terms generated from string theoretical effects can also be expressed using logarithmic functions \cite{solo1, solo2, solo4, solo5}. The logarithmic corrections terms for the entropy of a dilatonic black holes have been calculated \cite{jy}. Finally, the expansion of the partition function has also been used to calculate the corrections terms for the entropy of a black hole \cite{bss}. Such correction terms obtained by using the expansion of the partition function again are  logarithmic correction terms. The correction to the thermodynamics of black holes from generalized uncertainty principle has also been studied \cite{mi}. In this analysis the thermodynamics of the black holes gets modifies due to the generalization of the usual Heisenberg uncertainty principle.
It has been demonstrated this modified thermodynamics of the black holes predicts the existence
of a remnant for black holes. The existence of such remnants for black holes can have important phenomenological
consequences \cite{r1}.

As the quantum fluctuations can occur in all black hole geometries, we expect that the thermodynamics
of all black objects will get corrected due to thermal fluctuations. Thus, we can use the modified
relation between the entropy and area to analyze the corrections for the thermodynamics of any black object.
In this paper, we will analyze such correction terms for the thermodynamics of black Saturn.
The black Saturn are solutions to Einstein equations in higher dimensions. They are described by
a black hole surrounded by a black ring   \cite{0701035, 1007.3668}. This
 black ring is in thermodynamical equilibrium with  a spherical black hole.
 The thermodynamics of black Saturn has been studied
\cite{0705.3697}. The thermodynamic equilibrium is obtained because of the
rotation of the black ring. It is also possible to construct a black Saturn with
a  static black ring \cite{0802.0784, 0809.0154}. In this case, the the system remains in thermodynamic equilibrium because of an external magnetic field. It may be noted that
conditions for meta-stability of a black Saturn have also been studied
  \cite{0804.4575}. It has been demonstrated  that the black Saturn is  causal stably on the
closure of the domain of outer communications \cite{1102.3942}. The relation between the  black
Saturn and Myers-Perry black hole have also been analyzed \cite{1309.4414}. It may be noted that the thermodynamics of
a charged dilatonic black Saturn has also been studied \cite{1407.2009}.
It is expected that both the black hole and black ring in a black Saturn will reduce in size due to the Hawking radiation. Thus, at a certain stage quantum fluctuations in the geometry of a black Saturn will also become
important. To analyze the effect of these quantum fluctuations in the geometry of a black Saturn, we will need
to analyze the thermal fluctuations in the thermodynamics of black Saturn. So, we will study the corrections to the
thermodynamics of a black Saturn by considering thermal fluctuations around the equilibrium.

\section{Black Saturn}
In this section, we will review the thermodynamics of black Saturn.
The metric for black Saturn can be written as \cite{0701035}
\begin{eqnarray}\label{B1}
ds^{2}&=&-\frac{H_{y}}{H_{x}}\left[dt+(\frac{\omega_{\psi}}{H_{y}}+q)d\psi\right]^{2}\nonumber\\
&+&H_{x}\left[k^{2}P(d\rho^{2}+dz^{2})+\frac{G_{y}}{H_{y}}d\psi^{2}+\frac{G_{x}}{H_{x}}d\varphi^{2}\right],
\end{eqnarray}
where $q$ and $k$ are constants, and
\begin{eqnarray}\label{B2}
G_{x}&=&\frac{\mu_{4}}{\mu_{3}\mu_{5}}\rho^{2}\nonumber\\
G_{y}&=&\frac{\mu_{3}\mu_{5}}{\mu_{4}}.
\end{eqnarray}
Here we have used
\begin{equation}\label{B3}
P=(\mu_{3}\mu_{4}+\rho^{2})^{2}(\mu_{1}\mu_{5}+\rho^{2})(\mu_{4}\mu_{5}+\rho^{2}),
\end{equation}
and
\begin{equation}\label{B4}
\mu_{i}=\sqrt{\rho^{2}+(z-a_{i})^{2}}-(z-a_{i})=R_{i}-(z-a_{i}).
\end{equation}
The real constant parameters $a_{i}$ ($i=1,...,5$) satisfy the following condition,
\begin{equation}\label{B5}
a_{1}\leq a_{5}\leq a_{4}\leq a_{3}\leq a_{2}.
\end{equation}
Furthermore, we also have
\begin{eqnarray}\label{B6}
H_{x}&=&\frac{M_{0}+c_{1}^{2}M_{1}+c_{2}^{2}M_{2}+c_{1}c_{2}M_{3}+c_{1}^{2}c_{2}^{2}M_{4}}{F}\nonumber\\
H_{y}&=&\frac{1}{F}\frac{\mu_{3}}{\mu_{4}}
\left[\frac{\mu_{1}}{\mu_{2}}M_{0}-c_{1}^{2}M_{1}\frac{\rho^{2}}{\mu_{1}\mu_{2}}-c_{2}^{2}M_{2}
\frac{\mu_{1}\mu_{2}}{\rho^{2}}
+c_{1}c_{2}M_{3}+c_{1}^{2}c_{2}^{2}M_{4}\frac{\mu_{2}}{\mu_{1}}\right],
\end{eqnarray}
where $c_{1}$ and $c_{2}$ are real constants, and
\begin{eqnarray}\label{B7}
M_{0}&=&\mu_{2}\mu_{5}^{2}(\mu_{1}-\mu_{3})^{2}(\mu_{2}-\mu_{4})^{2}(\rho^{2}+
\mu_{1}\mu_{2})^{2}(\rho^{2}+\mu_{1}\mu_{4})^{2}(\rho^{2}+\mu_{2}\mu_{3})^{2},\nonumber\\
M_{1}&=&\mu_{1}^{2}\mu_{2}\mu_{3}\mu_{4}\mu_{5}\rho^{2}(\mu_{1}-\mu_{2})^{2}
(\mu_{2}-\mu_{4})^{2}(\mu_{1}-\mu_{5})^{2}(\rho^{2}+\mu_{2}\mu_{3})^{2},\nonumber\\
M_{2}&=&\mu_{2}\mu_{3}\mu_{4}\mu_{5}\rho^{2}(\mu_{1}-\mu_{2})^{2}
 (\mu_{1}-\mu_{3})^{2}(\rho^{2}+\mu_{1}\mu_{4})^{2}(\rho^{2}+\mu_{2}\mu_{5})^{2},\nonumber\\
M_{3}&=&2\mu_{1}\mu_{2}\mu_{3}\mu_{4}\mu_{5}(\mu_{1}-\mu_{3})(\mu_{1}-\mu_{5})
(\mu_{2}-\mu_{4})(\rho^{2}+\mu_{1}^{2})(\rho^{2}+\mu_{2}^{2})\nonumber \\ && \times
(\rho^{2}+\mu_{1}\mu_{4})(\rho^{2}+\mu_{2}\mu_{3})(\rho^{2}+\mu_{2}\mu_{5}),\nonumber\\
M_{4}&=&\mu_{1}^{2}\mu_{2}\mu_{3}^{2}\mu_{4}^{2}(\mu_{1}-\mu_{5})^{2}(\rho^{2}+\mu_{1}
\mu_{2})^{2}(\rho^{2}+\mu_{2}\mu_{5})^{2},\nonumber\\
\end{eqnarray}
with
\begin{eqnarray}\label{B8}
F&=&\mu_{1}\mu_{5}(\mu_{1}-\mu_{3})^{2}(\mu_{2}-\mu_{4})^{2}(\rho^{2}+\mu_{1}\mu_{3})\nonumber \\ && \times
(\rho^{2}+\mu_{2}\mu_{3})(\rho^{2}+\mu_{1}\mu_{4})(\rho^{2}+\mu_{2}\mu_{4})
(\rho^{2}+\mu_{2}\mu_{5})\nonumber\\
&&\times(\rho^{2}+\mu_{3}\mu_{5})(\rho^{2}+\mu_{1}^{2})(\rho^{2}+\mu_{2}^{2})(\rho^{2}+
\mu_{3}^{2})(\rho^{2}+\mu_{4}^{2})(\rho^{2}+\mu_{5}^{2}).
\end{eqnarray}
Here $\omega_{\psi}$ is expressed as
\begin{equation}\label{B9}
\omega_{\psi}=\frac{2}{F\sqrt{G_{x}}}\left[c_{1}R_{1}\sqrt{M_{0}M_{1}}-c_{2}R_{2}
\sqrt{M_{0}M_{2}}+c_{1}^{2}c_{2}R_{2}\sqrt{M_{1}M_{4}}-c_{1}c_{2}^{2}R_{1}\sqrt{M_{2}M_{4}}\right],
\end{equation}
where $R_{1}$ and $R_{2}$ given in the relation (\ref{B4}).
Free parameters of the model are fixed by \cite{1007.3668} as
\begin{equation}\label{B10}
L^{2}=a_{2}-a_{1},
\end{equation}
and
\begin{equation}\label{B11}
c_{1}=\pm\sqrt{\frac{2(a_{3}-a_{1})(a_{4}-a_{1})}{a_{5}-a_{1}}},
\end{equation}
We also have
\begin{equation}\label{B12}
c_{2}=\sqrt{2}(a_{4}-a_{2})\frac{\sqrt{(a_{1}-a_{3})(a_{4}-a_{2})(a_{2}-a_{5})(a_{3}-a_{5})}
\pm(a_{2}-a_{1})(a_{3}-a_{4})}
{\sqrt{(a_{1}-a_{4})(a_{2}-a_{4})(a_{1}-a_{5})(a_{2}-a_{5})(a_{3}-a_{5})}},
\end{equation}
and
\begin{equation}\label{B13}
k=\frac{2(a_{1}-a_{3})(a_{2}-a_{4})}{2(a_{1}-a_{3})(a_{2}-a_{4})+(a_{1}-a_{5})c_{1}c_{2}}=
\frac{2k_{1}\hat{k}_{2}}{2k_{1}\hat{k}_{2}+c_{1}c_{2}k_{3}},
\end{equation}
Here we have used $\hat{k}_{i}=1-k_{i}$ and,
\begin{equation}\label{B14}
k_{i}=\frac{a_{i+2}-a_{1}}{L^{2}},
\end{equation}
with $i=1,2,3$. The variable $q$ can be written as
\begin{equation}\label{B15}
q=\frac{2k_{1}c_{2}}{2k_{1}-2k_{1}k_{2}+c_{1}c_{2}k_{3}}.
\end{equation}
So, we can see that all parameters can written   in terms of   $a_{i}$ with $i=1,2,3,4,5$.

The Hawking temperatures for  the black saturn is obtained by combination
of the Hawking temperatures for the black hole with the  Hawking temperatures for the black ring.
This combined Hawking temperatures can be expressed as \cite{4a},
\begin{eqnarray}\label{T01}
T&=&\frac{1}{2\pi L}\sqrt{\frac{\hat{k}_{2}\hat{k}_{3}}{2\hat{k}_{1}}}\left(\frac{(1+k_{2})^{2}}
{1+\frac{k_{1}k_{2}\hat{k}_{2}\hat{k}_{3}}{k_{3}\hat{k}_{1}}c^{2}}\right)\nonumber\\
&+&\frac{1}{2\pi L}\sqrt{\frac{k_{1}\hat{k}_{3}(k_{1}-k_{3})}{2k_{2}(k_{2}-k_{3})}}
\left(\frac{(1+k_{2})^{2}}{1-(k_{1}-k_{2})c+\frac{k_{1}k_{2}\hat{k}_{1}}{k_{3}}c^{2}}\right),
\end{eqnarray}
where
\begin{equation}\label{T02}
c=\frac{1}{k_{2}}\left(\varepsilon\frac{k_{1}-k_{2}}{\sqrt{k_{1}\hat{k}_{2}\hat{k}_{3}(k_{1}-k_{3})}}-1\right).
\end{equation}
Here $\varepsilon=\pm1$, while $\varepsilon=0$  gives a naked singularity.
The entropy of the black Saturn, in absence of thermal fluctuations, can also be expressed as
\begin{eqnarray}\label{T03}
S_{0}&=&\frac{\pi^{2} L^{3}}{(1+k_{2})^{2}}\sqrt{\frac{2\hat{k}_{1}^{3}}{\hat{k}_{2}\hat{k}_{3}}}
\left(1+\frac{k_{1}k_{2}\hat{k}_{2}\hat{k}_{3}c^{2}}{k_{3}\hat{k}_{1}}\right)\nonumber\\
&+&\frac{\pi^{2} L^{3}}{(1+k_{2})^{2}}\sqrt{\frac{2k_{2}(k_{2}-k_{3})^{3}}{k_{1}(k_{1}-k_{3})
\hat{k}_{3}}}\left(1-(k_{1}-k_{2})c+\frac{k_{1}k_{2}\hat{k}_{3}c^{2}}{k_{3}}\right),
\end{eqnarray}
This is the entropy of the combination of the black hole and black ring.
The ADM mass of black Saturn, which can interpreted as enthalpy $H=M_{ADM} $,  is given by    \cite{4a,Dolan1,JJP},
\begin{eqnarray}\label{T04}
 M_{ADM}=\frac{3\pi L^{2}}{4k_{3}(1+k_{2}c)^{2}}\left(k_{3}
(\hat{k}_{1}+k_{2})-2k_{2}k_{3}(k_{1}-k_{2})c+k_{2}[k_{1}-k_{2}k_{3}(\hat{k}_{2}+k_{1})]c^{2}\right).
\end{eqnarray}
We will use this expression to study enthalpy and therefore $PV$ diagram.

\section{Corrected Thermodynamics}
In this section, we will analyze the corrections to the thermodynamics of a black Saturn because
of thermal fluctuations. These thermal fluctuations will become important as the black Saturn reduces in size
due to the Hawking radiation.
We can now write the partition function for this system  as
\begin{equation}
 Z = \int D g  D A \exp (- I),
\end{equation}
where $I \to -i I$ is the Euclidean action for this system. This partition function can also be written as
 \cite{l1, SPR}
\begin{equation}
 Z = \int_0^{\infty} dE \, \,  \rho (E) e^{-\beta E},
\end{equation}
where $\beta$ is the inverse of the temperature.
We can write an expression for the density of states as
\begin{eqnarray}
 \rho (E) = \frac{1}{2 \pi i} \int_{c- i\infty}^{c + i \infty} d \beta \, \, e ^{S(\beta)} ,
\end{eqnarray}
where
\begin{equation}
 S = \beta  E   + \ln Z.
\end{equation}
The quantum fluctuations in the geometry of space-time, will lead to the thermal fluctuations in the thermodynamics
of black Saturn. Thus, if the $S(\beta)$ is inverse of the corrected temperature, then we can expand it around the equilibrium temperature $\beta_0$,
\begin{equation}
 S = S_0 + \frac{1}{2}(\beta - \beta_0)^2
 \left(\frac{\partial^2 S(\beta)}{\partial \beta^2 }\right)_{\beta = \beta_0} + \cdots .
\end{equation}
Neglecting the higher order corrections, we obtain
\begin{eqnarray}
 \rho (E) = \frac{e^{S_0}}{ 2 \pi i}  \int_{c- i\infty}^{c + i \infty}  d \beta \, \,  \exp \left( \frac{1}{2}
 (\beta- \beta_0)^2 \left(\frac{\partial^2 S(\beta)}{\partial \beta^2 }\right)_{\beta = \beta_0}   \right).
\end{eqnarray}
We can write this expression as
\begin{equation}
  \rho(E) = \frac{e ^{S_{0}}}{\sqrt{2 \pi }} \left[\left(\frac{\partial^2 S(\beta)}{\partial
  \beta^2 }\right)_{\beta = \beta_0}\right]^{-1/2}.
\end{equation}
Now, we obtain
\begin{equation}
  S = S_0 -\frac{1}{2}
  \ln \left[\left(\frac{\partial^2 S(\beta)}{\partial \beta^2 }\right)_{\beta = \beta_0}\right]^{1/2}.
\end{equation}
It may be noted that this second derivative of entropy is actually a fluctuation squared of the energy, so
we can write   \cite{l1, SPR}
\begin{equation}
 \left[\left(\frac{\partial^2 S(\beta)}{\partial \beta^2 }\right)_{\beta = \beta_0}\right]^{1/2} = C_0 T^2
\end{equation}
Hence,
\begin{equation}
S = S_0 -\frac{1}{2} \ln C_0T^2.
\end{equation}
Here, $S_0$ is the original entropy of the combination of the black hole and the black ring given by the equation (\ref{T03}).
Thus, the entropy of both black ring and black hole will get corrections to it because of thermal fluctuation.
Now, we will introduce a variable $\alpha$ to parameterize the effect of thermal fluctuations on the thermodynamics
of black Saturn. Thus, we will write the expression for the entropy as
\begin{equation}\label{L1}
S = S_{0}-\frac{\alpha}{2} \ln{C_{0}T^{2}},
\end{equation}
where
\begin{equation}\label{L2}
C_{0}=T\frac{\partial S_{0}}{\partial T},
\end{equation}

\begin{figure}[th]
 \begin{center}
 \includegraphics[scale=.20]{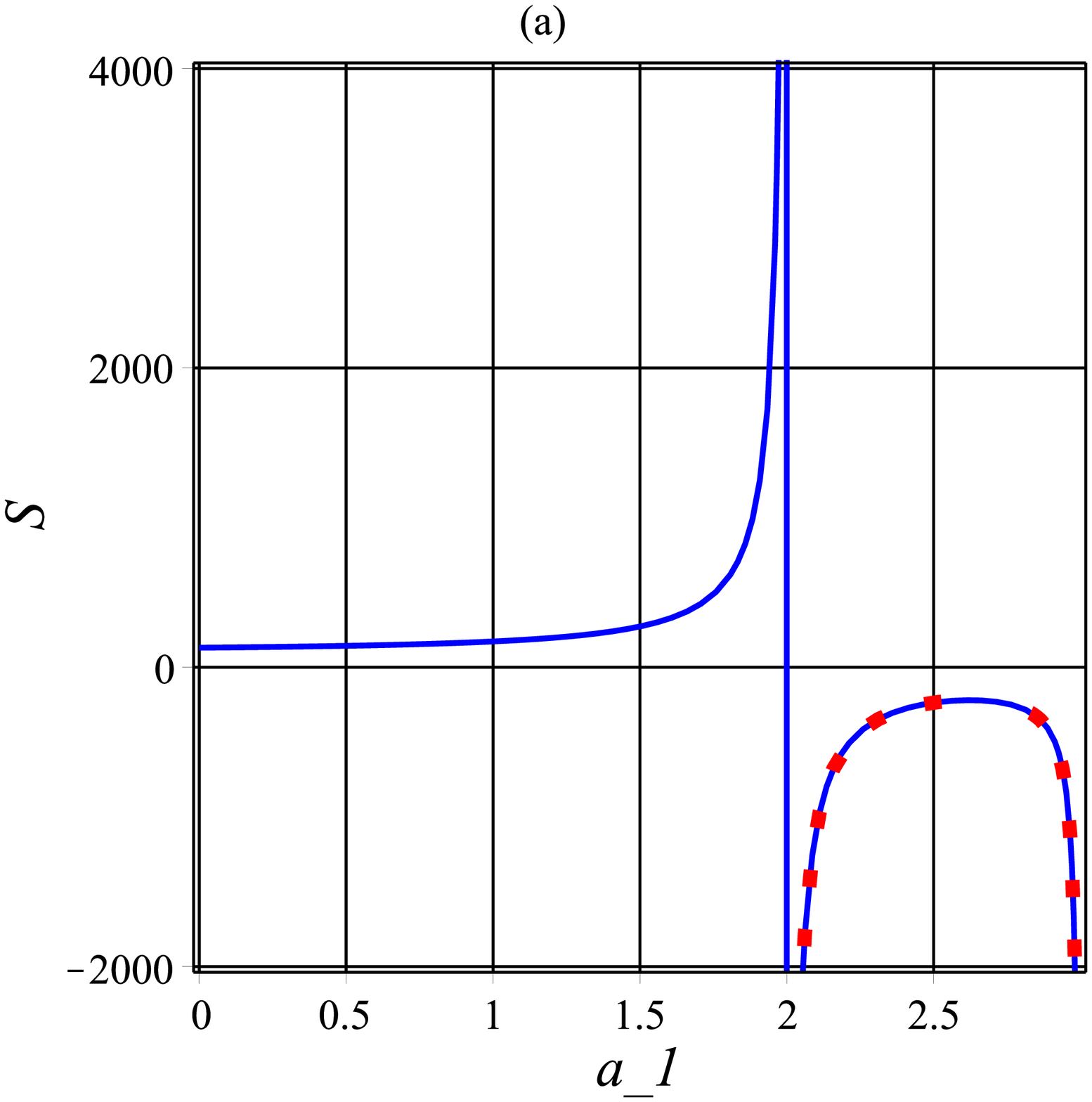}\includegraphics[scale=.20]{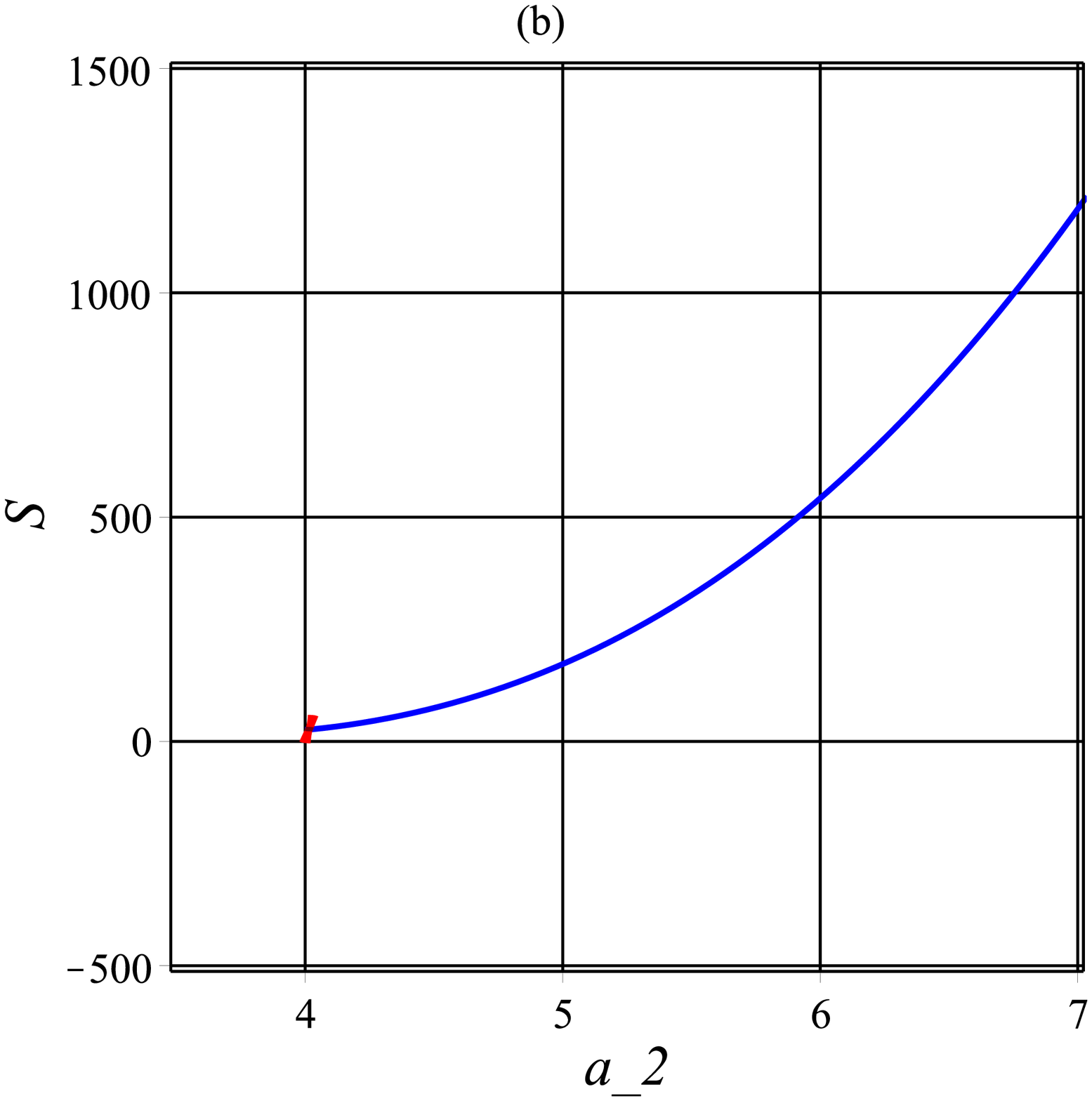}\\
 \includegraphics[scale=.20]{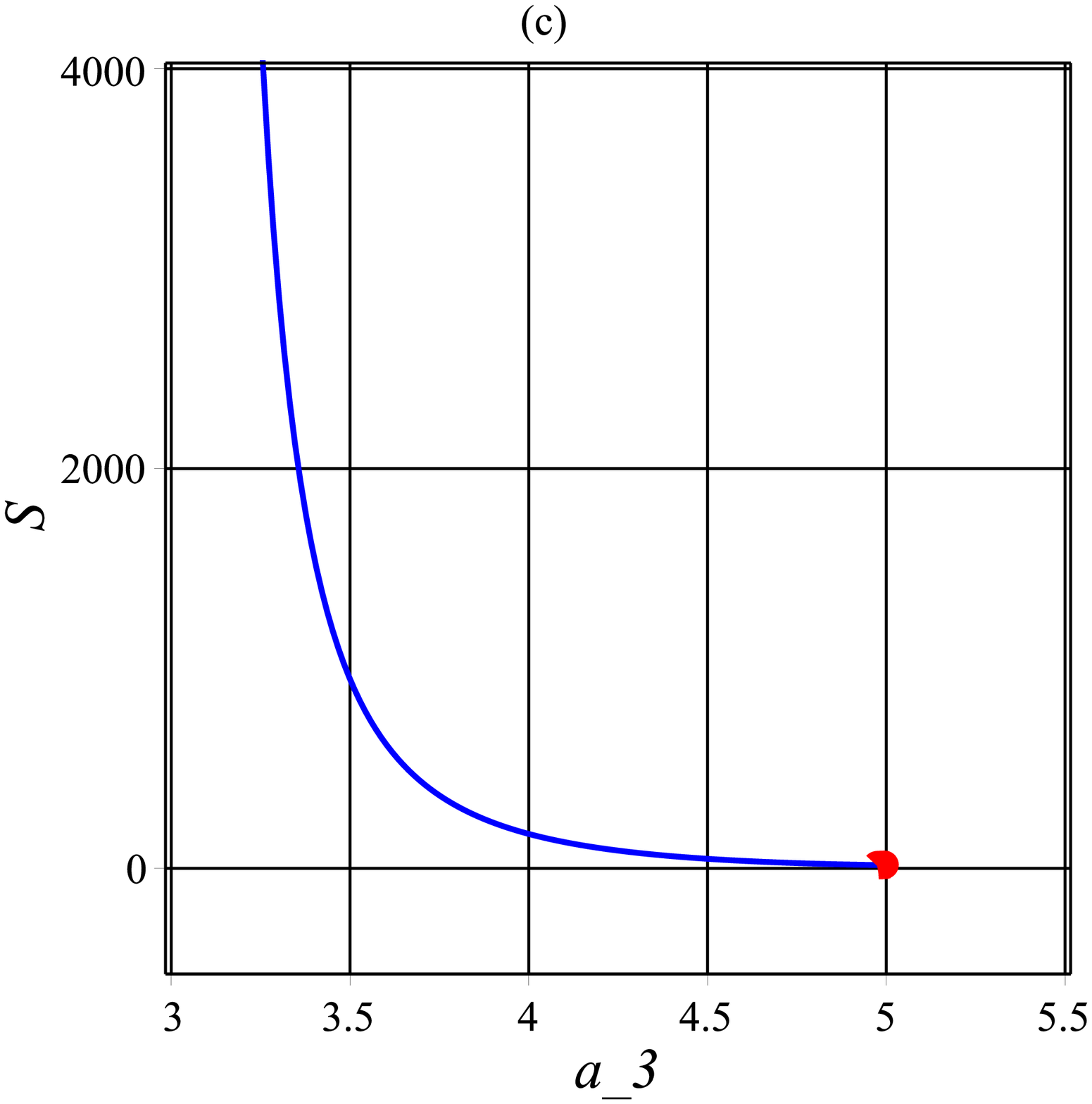}\includegraphics[scale=.20]{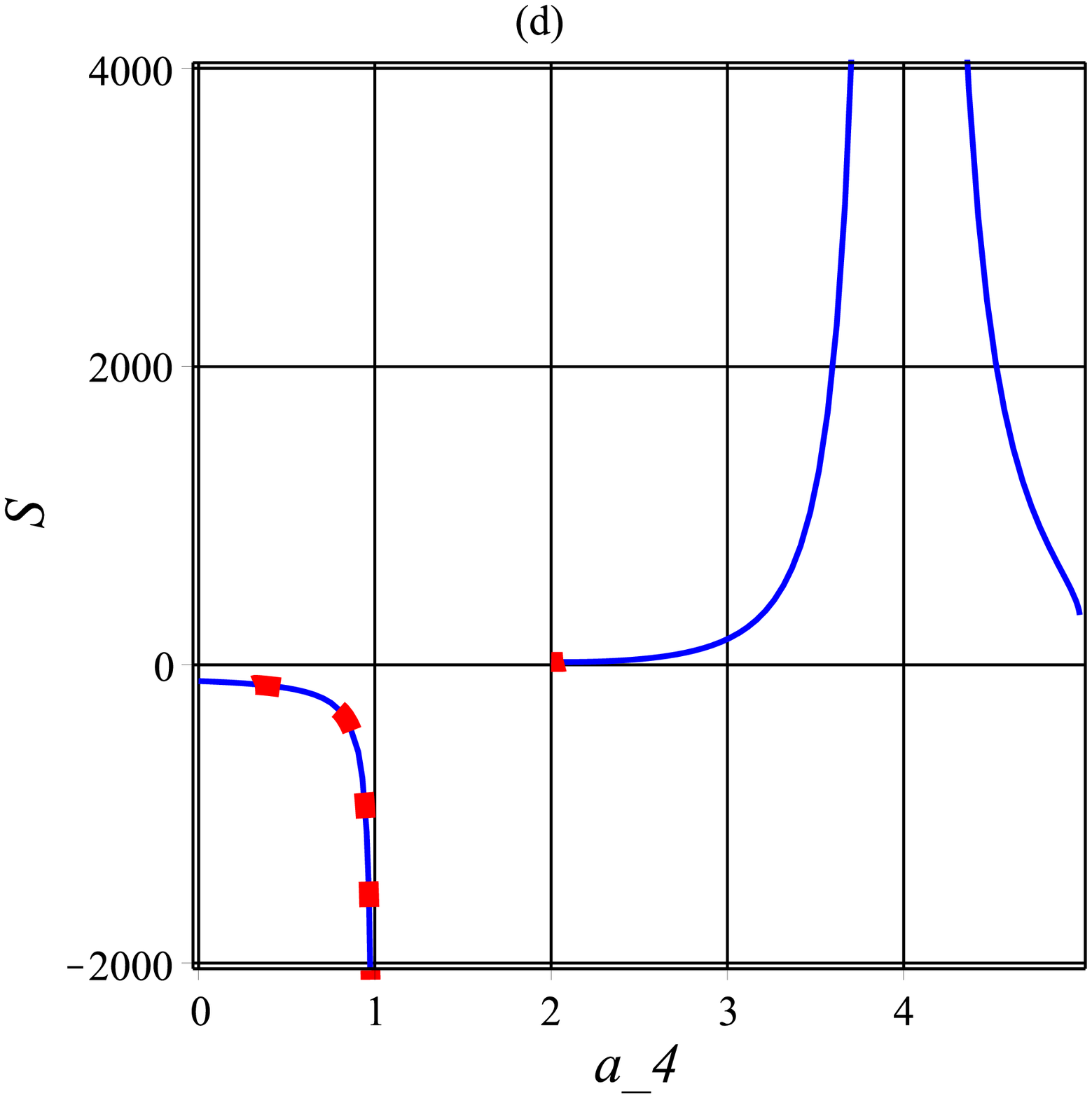}\\
 \includegraphics[scale=.20]{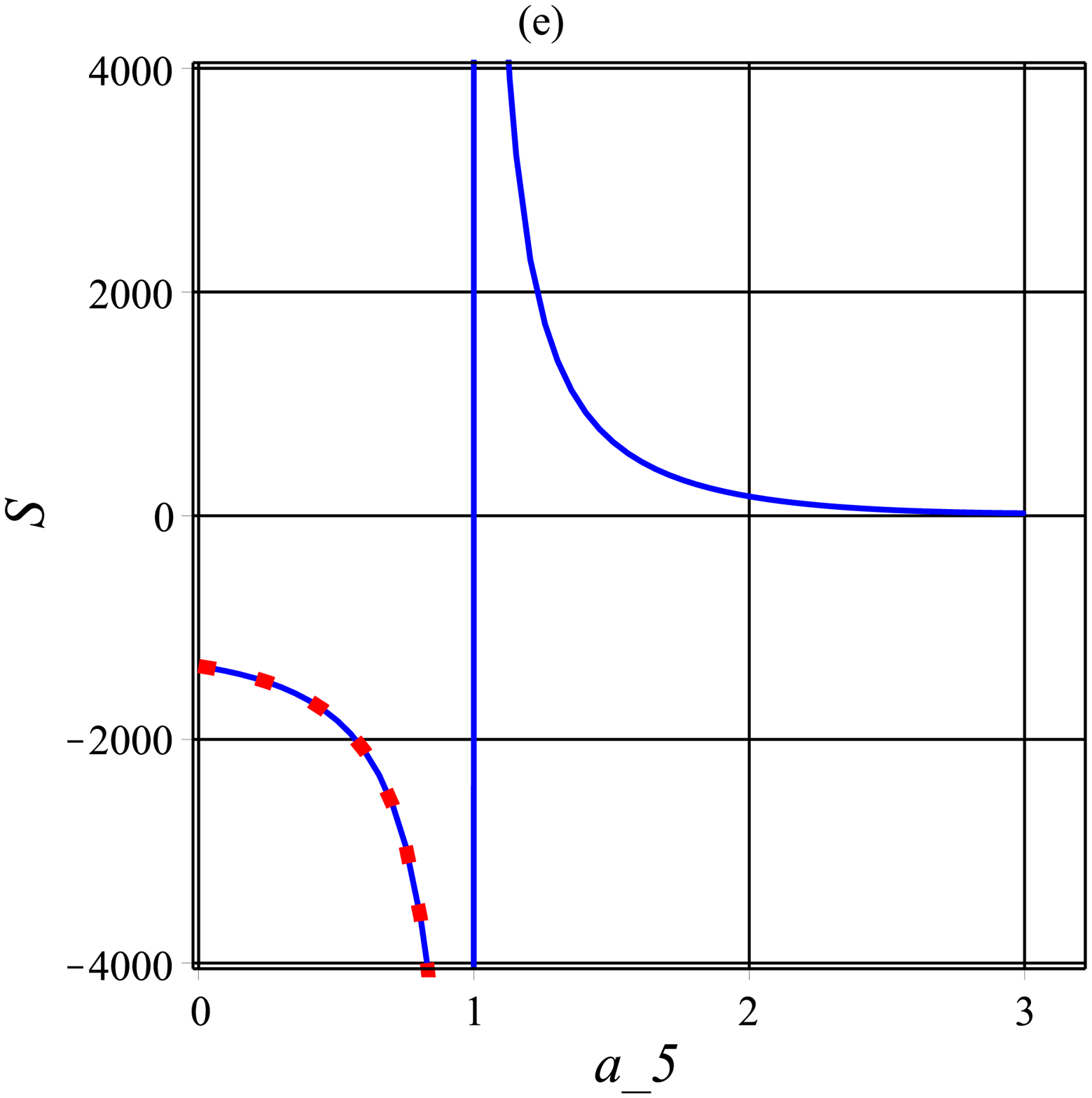}
 \caption{Entropy in terms of $a_{i}$, solid blue lines denote the case of $\alpha=0$ (ordinary entropy) and dashed red lines denote the case of $\alpha=1$ (logarithmic corrected entropy) (a) $a_{2}=5$, $a_{3}=4$, $a_{4}=3$ and $a_{5}=2$, (b) $a_{1}=1$, $a_{3}=4$, $a_{4}=3$ and $a_{5}=2$, (c) $a_{1}=1$, $a_{2}=5$, $a_{4}=3$ and $a_{5}=2$, (d) $a_{1}=1$, $a_{2}=5$, $a_{3}=4$ and $a_{5}=2$, (e) $a_{1}=1$, $a_{2}=5$, $a_{3}=4$ and $a_{4}=3$.}
 \label{fig1}
 \end{center}
 \end{figure}

We can infer from relation (\ref{B14})   that all thermodynamics quantities are depend on
$a_{1}$, $a_{2}$, $a_{3}$, $a_{4}$ and $a_{5}$. Therefore, we can calculate
thermodynamics quantities in terms of $a_{i}$ with $i=1,...,5$.
First of all, for simplicity we can fix four of them according to condition
given by the equation (\ref{B5}), and obtain thermodynamics quantities in
terms of only one free parameter. Therefore we will consider five different cases.
Using the equations (\ref{T01}), (\ref{T03}) and (\ref{L1}), we can study
logarithmic corrected entropy. For the five   cases of fixed parameters, we
can analyze the behavior of entropy in the plots of Fig. \ref{fig1}. We fix parameters as
$a_{1}=1$, $a_{2}=5$, $a_{3}=4$, $a_{4}=3$ and $a_{5}=2$, which satisfy condition (\ref{B5}).

In Fig. \ref{fig1} (a), we vary $a_{1}$ with the condition $a_{1}\leq2$, and in
Fig. \ref{fig1} (e), we vary $a_{5}$, with the condition $1\leq a_{5}\leq3$. It is observed that both are in agreement
with the condition (\ref{B5}). We can see that both solutions are only valid in absence
of logarithmic correction, so we are not allowed to use fixed parameters as illustrated
by Fig. \ref{fig1} (a) and (e). In Fig. \ref{fig1} (b), we vary $a_{2}$ with the condition $a_{2}\geq4$
and observe  that presence of logarithmic correction fix $a_{2}=a_{3}$. Similar result obtained by Fig.
\ref{fig1} (c). Here we vary $a_{3}$ with the condition $a_{3}\leq5$, and see that presence of logarithmic
correction fix $a_{3}=a_{2}$. In Fig. \ref{fig1} (d), we vary $a_{4}$ with the condition $2\leq a_{4}\leq4$,
and observe that it is only possible to consider the case of $a_{4}=a_{5}$
in presence of logarithmic corrections. Therefore, the
logarithmic correction fix free parameters $(a_{1}, a_{2}, a_{3}, a_{4}, a_{5})$.  Thus, we can consider the following
  example   $(1, 4, 4, 3, 2)$,
$(1, 5, 5, 3, 2)$ or $(1, 5, 4, 2, 2)
$. It implies  that if we fix four parameter, then the last parameter also should be fixed.
When, all parameters fixed then, temperature and entropy and all thermodynamics quantities
are constants while we need them as a function of parameters.

\begin{figure}[th]
 \begin{center}
 \includegraphics[scale=.20]{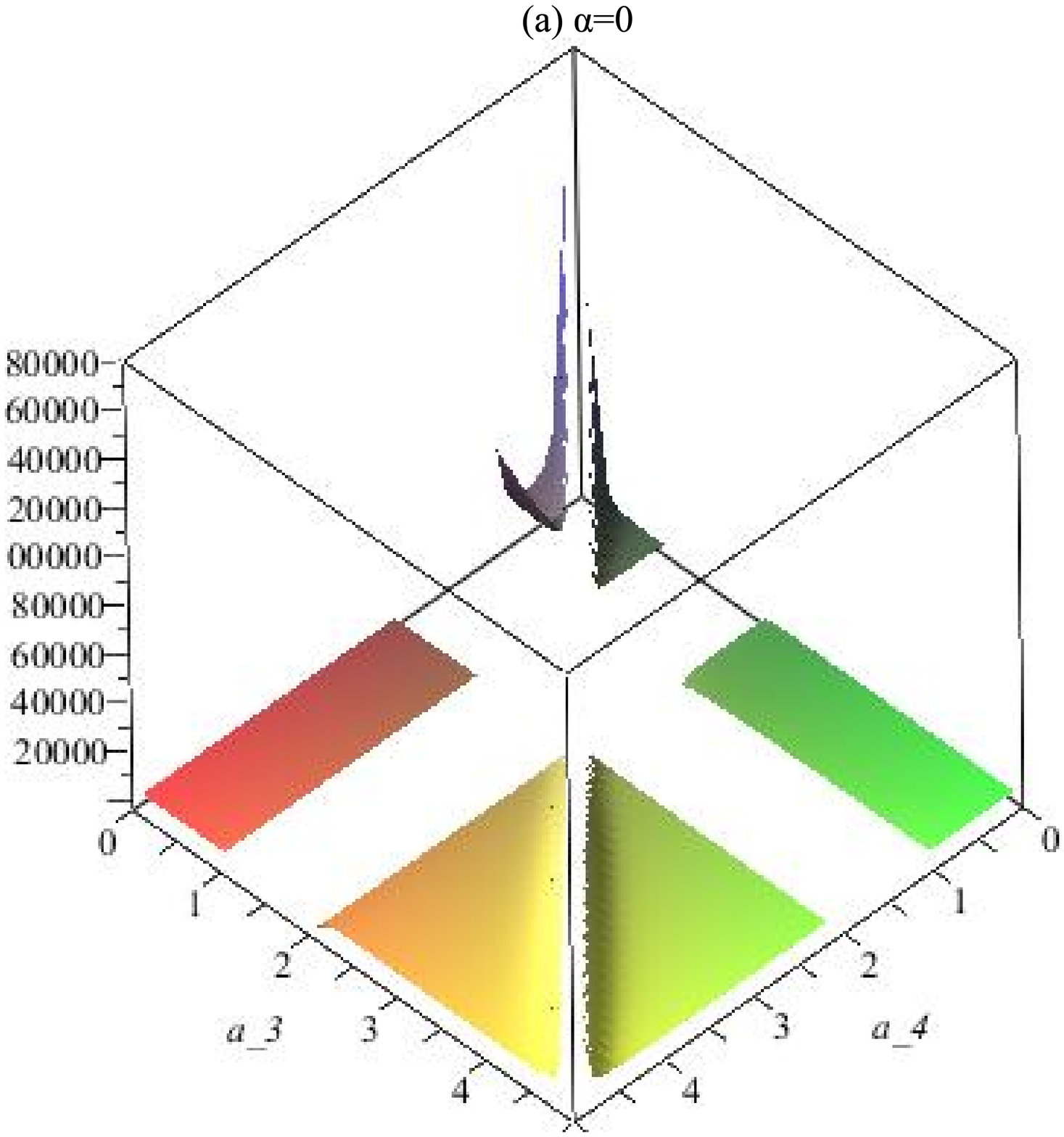}\includegraphics[scale=.23]{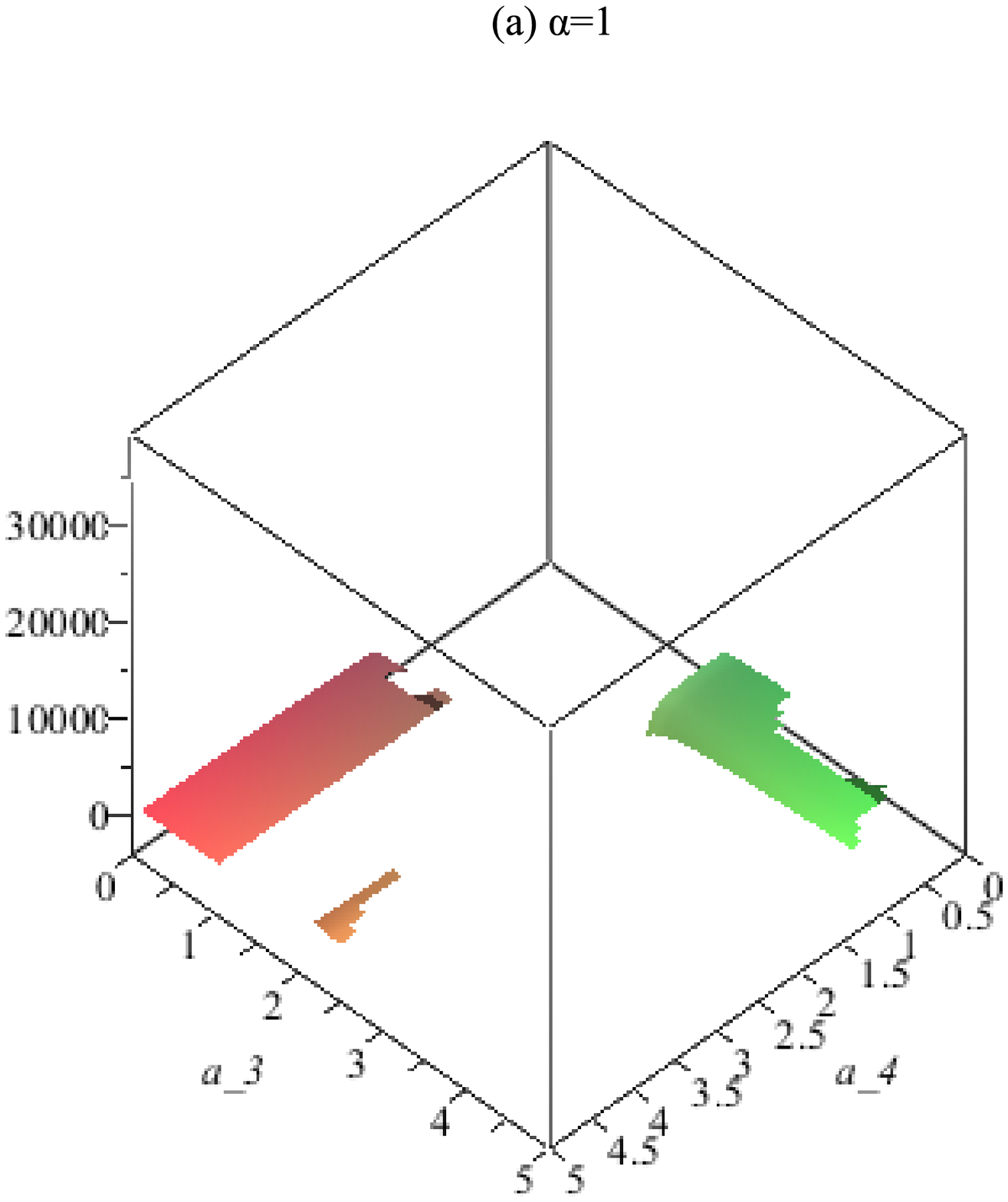}\\
 \includegraphics[scale=.20]{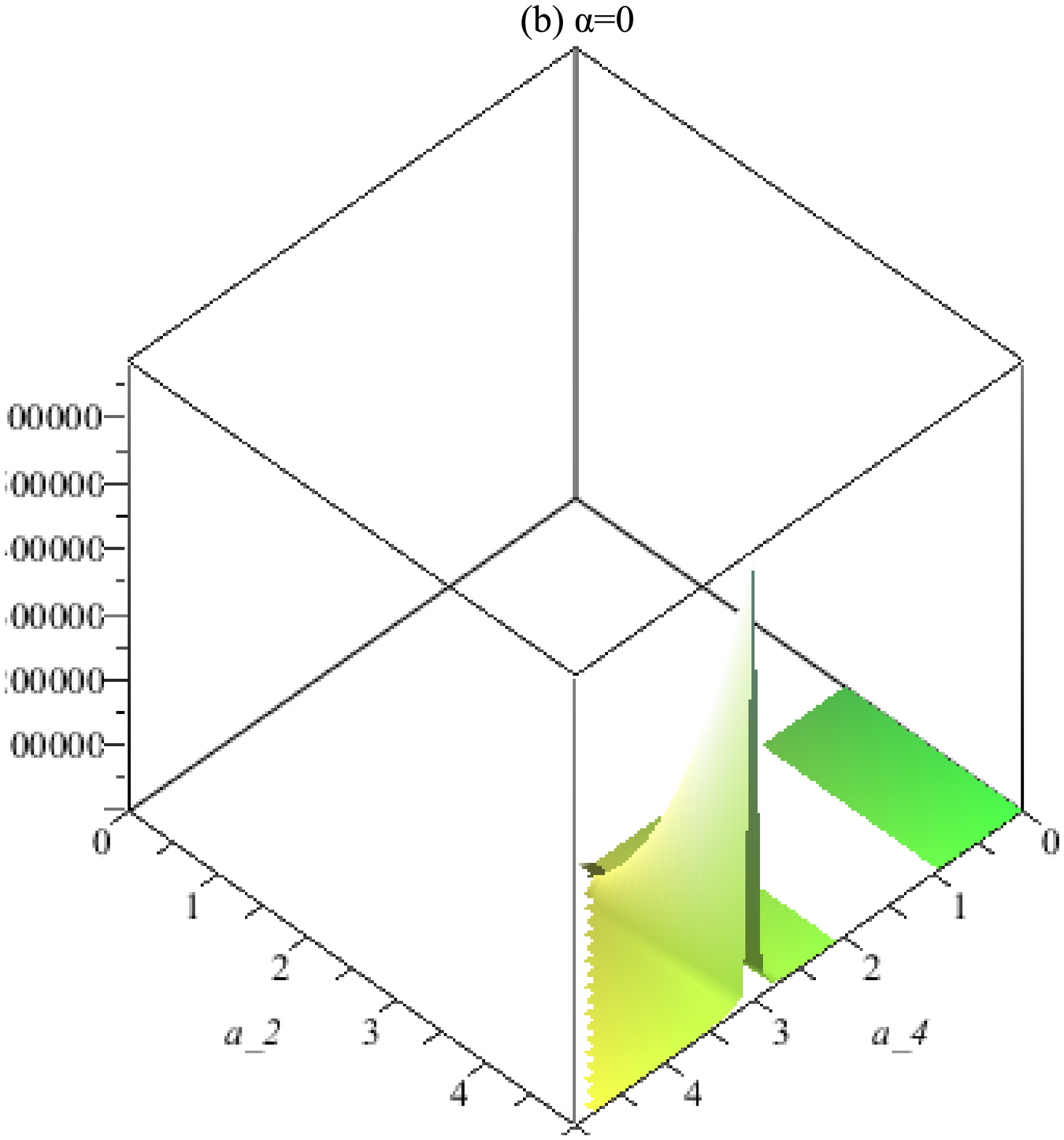}\includegraphics[scale=.20]{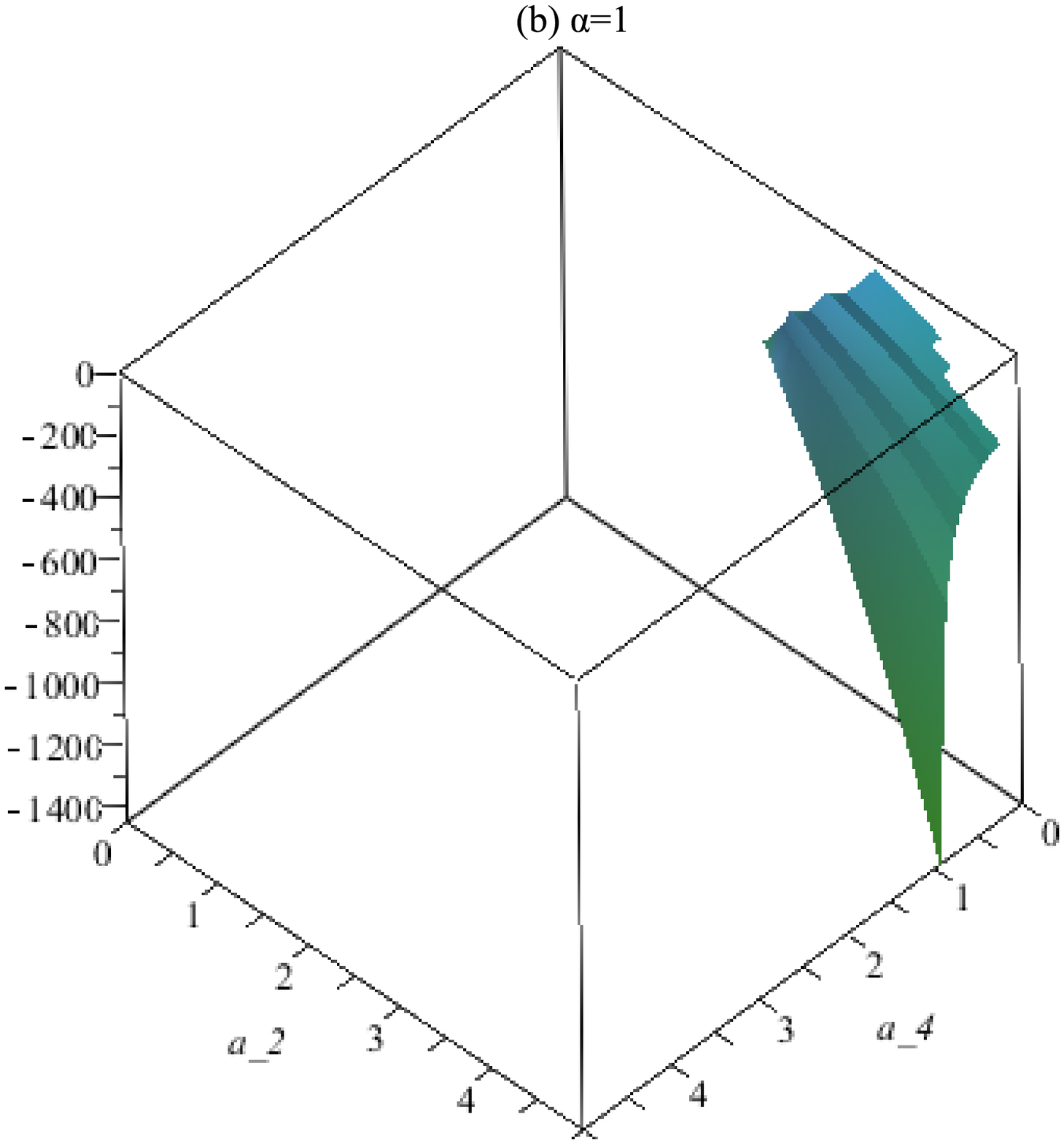}\\
 \includegraphics[scale=.20]{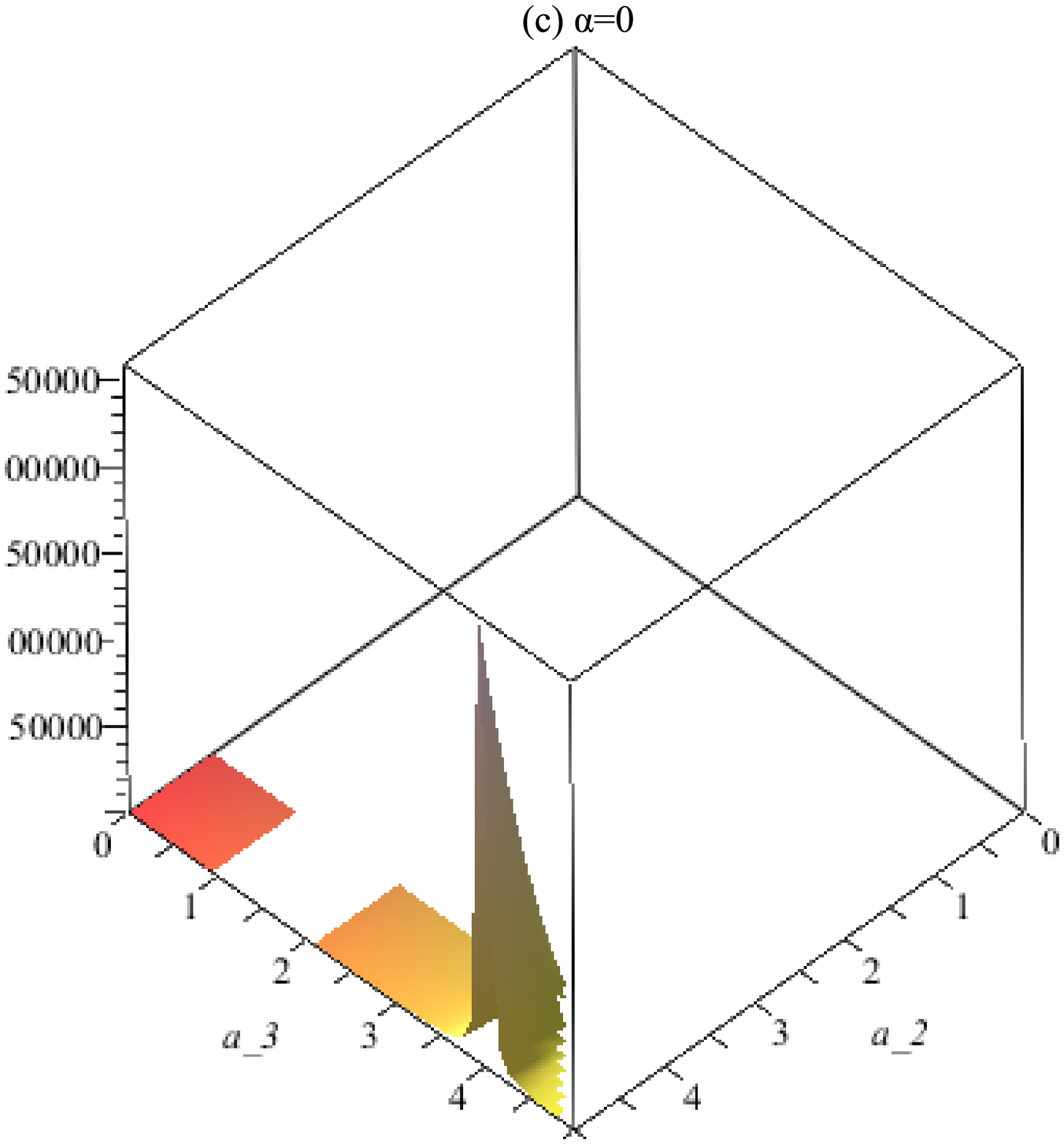}\includegraphics[scale=.20]{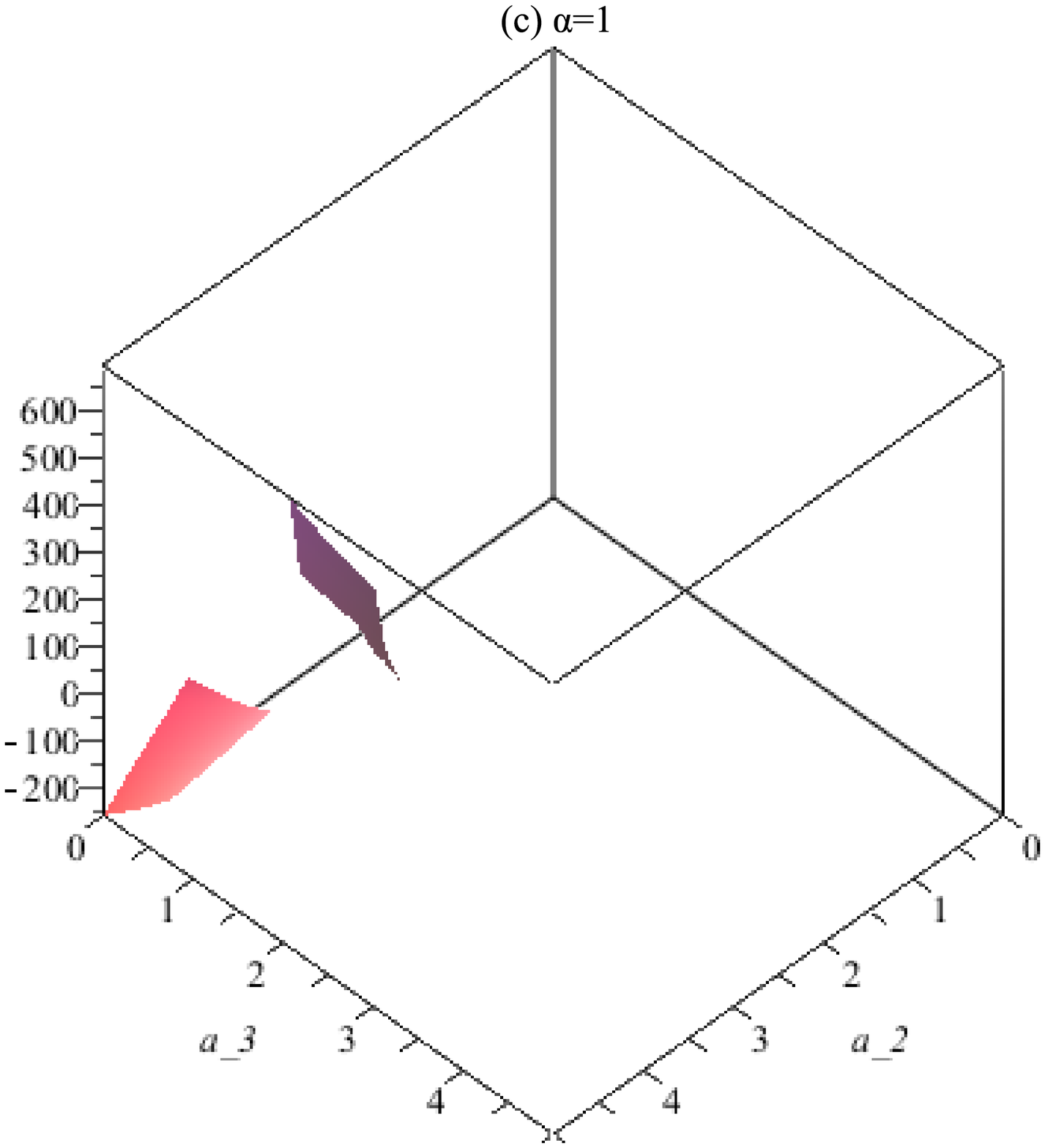}
 \caption{Entropy in terms of $a_{2}$, $a_{3}$ and $a_{4}$ with $a_{1}=1$ and $a_{5}=2$.
 Left plots denote the case of $\alpha=0$ (ordinary entropy) and right plots denote the
 case of $\alpha=1$ (logarithmic corrected entropy) (a) $a_{2}=5$, (b) $a_{3}=3$, (c) $a_{4}=4$.}
 \label{fig2}
 \end{center}
 \end{figure}

So, we will only fix three of these parameters and keep two of them as variable.
From Fig. \ref{fig1}, we observe  that by considering $a_{1}$ and $a_{5}$ as variable, we do not obtain a
  physical results. Therefore, we will fix $a_{1}=1$ and $a_{5}=2$, and vary other parameters
(according to the condition (\ref{B5})). In the  plots of the Fig. \ref{fig2}, we plot the
entropy corrected by the logarithmic term  (\ref{L1}), and ordinary entropy (\ref{T03}) as a function of
two $a_{i}$. In plots of (a), we fix $a_{2}$ and vary $a_{3}$ and $a_{4}$. This is done  for both
$\alpha=0$ and $\alpha=1$. In plots of (b), we fix $a_{3}$ and vary $a_{2}$ and $a_{4}$. This is again done for
  $\alpha=0$ and $\alpha=1$. In plots of (c), we fix $a_{4}$ and vary $a_{3}$
along with $a_{2}$. Here again we consider  both the values of $\alpha$, i.e.,
$\alpha=0$ and $\alpha=1$. The
   allowed values of $a_{3}$ and $a_{4}$ can be  inferred from the plot of Fig. \ref{fig2} (a).
   This is done by considering those values that   satisfy the condition (\ref{B5}). Thus, we consider
  $a_{3}\geq2$ and $a_{4}\geq a_{3}$, and   observe  that there is no entropy in
this region. The same fact can be obtained by using
  plot of Fig. \ref{fig2} (c). This is done by fixing $a_{4}$ and varying $a_{3}$ and $a_{2}$ according the
condition (\ref{B5}). Here we again obtain nothing. Therefore the only choice we have is to   selected values like
Fig. \ref{fig2} (b). However, for these values the   the entropy is negative.

\begin{figure}[th]
 \begin{center}
 \includegraphics[scale=.3]{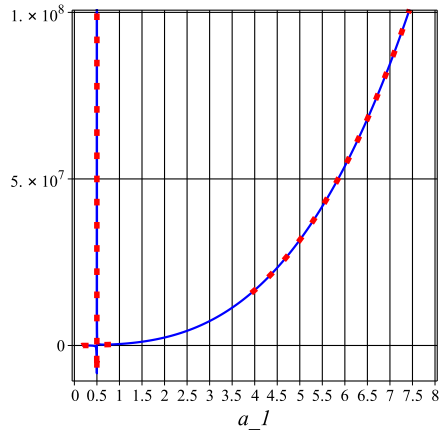}
 \caption{Entropy in terms of $a_{1}$, with $\alpha=0$ (solid blue) and $\alpha=1$ (dotted red).}
 \label{fig3}
 \end{center}
 \end{figure}

It we obtain the   third solution. We can   relate all parameter to each other and rewrite
them in terms of only one free parameter as a variable.  Therefore all of them are
considered as variables. Even though there are   several choices,   we will consider the     following choices,
\begin{eqnarray}\label{C0}
a_{2}&=&5a_{3}\nonumber\\
a_{3}&=&4a_{4}\nonumber\\
a_{4}&=&3a_{5}\nonumber\\
a_{5}&=&2a_{1}^{2}.
\end{eqnarray}
Therefore the only free parameter of model is $a_{1}$. Using ansatz (\ref{C0}) in the
corrected entropy (\ref{L1}) give us results which  is illustrated by the Fig. \ref{fig3}.
As  was expected, the parameters for black hole ($a_{1}$ and hence other parameters) get restricted
in presence of logarithmic corrections ($\alpha=1$). The values we  selected  can be used to obtain, $a_{1}\geq4$
and   $a_{5}\geq32$, $a_{4}\geq66$, $a_{3}\geq264$ and $a_{2}\geq1320$. On the other hand, it is also possible
to consider
\begin{eqnarray}\label{C00}
0.1<a_{1}<0.5,\nonumber\\
0.5<a_{1}<0.9.
\end{eqnarray}
There is a singularity at $a_{1}=0.5$. It is possible to  use specific heat for
 choosing one of above mentioned solutions.
It is possible to choose other possibilities and obtain similar results.

\begin{figure}[th]
 \begin{center}
 \includegraphics[scale=.3]{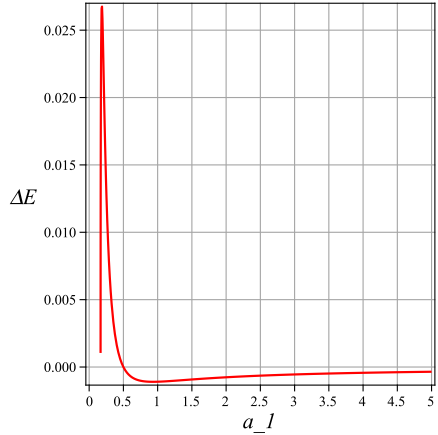}
 \caption{Change of internal energy (due to logarithmic correction) in terms of $a_{1}$.}
 \label{fig4}
 \end{center}
 \end{figure}

It is possible to demonstrate that
\begin{equation}\label{C1}
\frac{\partial S}{\partial a_{1}}=\frac{\partial S_{0}}{\partial a_{1}}-\frac{\alpha}{2}
(1+\frac{1}{T})\frac{\partial T}{\partial a_{1}}.
\end{equation}
Thus, we can write the  internal energy as,
\begin{equation}\label{C2}
E=\int{T dS}=E_{0}-\frac{\alpha}{2}T-\frac{\alpha}{4}T^{2},
\end{equation}
where
\begin{equation}\label{C3}
E_{0}=\int{T dS_{0}}.
\end{equation}
So we can study behavior of $\Delta E=E-E_{0}$ in terms of $T$, and this in turn can be expressed  in terms of
$a_{1}$ (see Fig. \ref{fig4}). There are two special choices of $a_{1}$, where the
  logarithmic correction do not contribute to the result. The first one being
  $a_{1}=0.5$, and this   corresponds to a singularity in
the entropy (see Fig. \ref{fig3}), and $a_{1}\approx20$ and this  is not in appropriate domain
of $0.1< a_{1}<0.5$ and $0.5<a_{1}<0.9$. This    will be demonstrated to be a   requirement of stability. So,
we can infer that, in the case of $0.1< a_{1}<0.5$ corrected energy is bigger than ordinary energy
($E>E_{0}$). On the other hand, for the case of $a_{1}>0.5$,  we have $E<E_{0}$.

\begin{figure}[th]
 \begin{center}
 \includegraphics[scale=.3]{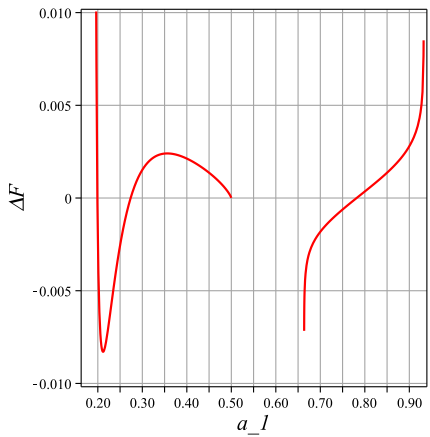}
 \caption{Change of Helmholtz free energy (due to logarithmic correction) in terms of $a_{1}$.}
 \label{fig5}
 \end{center}
 \end{figure}

Then, using the relation (\ref{T04}) and,
\begin{equation}\label{C4}
H=E+PV,
\end{equation}
we can investigate $PV$ diagram. We can also  analyze the behavior of Gibbs free energy  because
\begin{equation}\label{C5}
G=F+PV,
\end{equation}
where
\begin{equation}\label{C6}
F=E-TS,
\end{equation}
is Helmholtz free energy. It is easy to find that,
\begin{equation}\label{C7}
F=F_{0}-\frac{\alpha}{2}T(1-\ln{C_{0}T^{2}})-\frac{\alpha}{4}T^{2},
\end{equation}
where
\begin{equation}\label{C8}
F_{0}=E_{0}-TS_{0}.
\end{equation}
So, we can study $\Delta F=F-F_{0}$ in terms of $a_{1}$ (see Fig. \ref{fig5}).
Fig. \ref{fig5} shows that the Helmholtz free energy   exists only for selected values of $a_{1}$ as,
\begin{eqnarray}\label{C9}
0.2\leq a_{1}<0.5,\nonumber\\
0.65<a_{1}<1,
\end{eqnarray}
This is more restricted than previous result. It is clear that there are
  different values of $a_{1}$, where $F=F_{0}$, such that   the effects from the logarithmic corrections are
canceled by effect of $a_{1}$. In this case, we have  $a_{1}=0.2, 0.275, 0.5, 0.78$.
It may be noted that   bound on the  parameters of a black saturn have been obtained  from the existence of Helmholtz free energy. 
We will demonstrate this to be related to the a stability condition, i.e., it is related to the positivity of the heat capacity. 
Hence, the black saturns are stable only for certain values of parameters. Furthermore, we would like to clarify that it has been demonstrated 
that black saturns are generally unstable. This is because   higher entropy solution with the same charges always exists. 
However, it is possible for  black Saturns to be  metastable. We will use the metastability of black saturns to find bounds on 
the parameters. So, basically we analyse the bounds from metastability of black saturns, and also the effect of thermal fluctuations on such 
bounds.  

\section{Phase Transition}
\begin{figure}[th]
 \begin{center}
 \includegraphics[scale=.3]{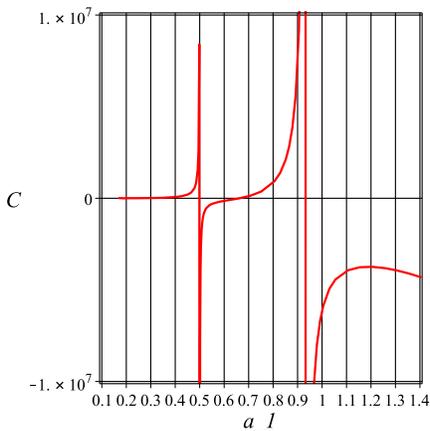}
 \caption{Specific heat in terms of $a_{1}$.}
 \label{fig6}
 \end{center}
 \end{figure}

In this section, we will analyze the phase transition in the black Saturn.
Using the sign of specific heat at constant volume,
\begin{equation}\label{P01}
C=T\frac{\partial S}{\partial T},
\end{equation}
one can investigate phase transition of black hole. We will analyze the instability in
presence of logarithmic corrections.
Using the relation (\ref{C1}), above equation can be written as
\begin{equation}\label{P02}
C=C_{0}-\frac{\alpha}{2}(1+T).
\end{equation}
It is easy to check that change of specific heat due to logarithmic correction is small as comparison to $10^7$. Therefore, we can observe  that logarithmic
corrections does not  effect the phase transition and thermodynamics stability of black Saturn.
As these logarithmic correction are generated from the thermal fluctuations, which in turn are generated from
quantum corrections, we can infer that the black Saturn remain stable even in presence of quantum corrections.
Thus, the black Saturn continue to remain stable as they get smaller due to the Hawking radiation.
In the Fig. \ref{fig6}, we have obtained phase transition points, which are $a_{1}=0.5, 0.65, 0.92$. Thus, there are two choices for black Saturn parameter
\begin{eqnarray}\label{P03}
0.2\leq a_{1}\leq0.5,\nonumber\\
0.65 \leq a_{1} \leq 0.9.
\end{eqnarray}
The black Saturn has thermodynamics stability for both these choices.
There are two singular points, $a_{1}=0.5 $ and $a_{1}\approx0.92$. Therefore, we can find two
intervals for possible values of $a_{1}$. This is given by  overlap of the regions given by (\ref{C00}) and (\ref{C9}). Now, the equation (\ref{P03}) can be written as
\begin{equation}\label{P04}
0.2\leq a_{1}<0.9, \hspace{2cm} a_{1}\neq0.5.
\end{equation}
However there is a condition given by (\ref{B5}). This condition uses (\ref{C0}) to infer that
$a_{5}=2a_{1}^{2}$ and $a_{5}\geq a_{1}$. Both  these conditions are satisfied if
\begin{equation}\label{P05}
0.6\leq a_{1}<0.9.
\end{equation}
Thus, we have been able to analyze the effect of thermal corrections on the stability of black Saturn.

\section{Conclusions}
In this paper,  we studied thermodynamics quantities of a black Saturn. We were able to analyze the effect of thermal fluctuations to the thermodynamics of black Saturn. The leading order corrections term to the entropy of black Saturn were the standard logarithmic corrections. It was not possible to fix all the free  parameters for the black Saturn. However, we were able to use the stability condition to obtain a bound for these free parameters. The bound we obtain for these parameters could be expressed as
 $
0.6  \leq  a_{1}  < 0.9,\,\,
0.72 \leq  a_{5}  < 1.62,\,\,
2.16 \leq  a_{4}  < 4.86,\,\,
8.64 \leq a_{3}  < 19.44,\,\,
43.2 \leq a_{2}  < 97.2.
$
We also analyzed the phase transition for the black Saturn. We were able to explicitly calculate the
points were phase transition can take place. It was demonstrated that the thermal fluctuations
do not effect stability of the black Saturn, and so their effect can be neglected when analyzing the
phase transition in the black Saturn. It will be interesting to explicitly calculate the partition function
for density of states of the black Saturn. This can then be used to calculate the Helmholtz
free energy and entropy for the black Saturn. Then, we can compare the results obtained to
the analysis done in this paper.  The thermodynamics of
a charged dilatonic black Saturn has also been studied \cite{1407.2009}.
In this analysis a
charged black rings along with
a black Saturn has been studied using the   Einstein-Maxwell-dilaton theory.
This analysis was performed in five dimensions. This was done by embedding
a neutral black ring and black Saturn solutions in six dimensions. They were then boosted with respect to the time coordinate and the sixth dimension. Then, the  Kaluza-Klein reduction was used to obtain the  charged solutions. The phase diagram was also studied for this system. It would be interesting to analyze the effect of thermal fluctuations on the thermodynamics of this charged dilatonic black Saturn. The entropy of this charged charged dilatonic black Saturn is also expected to get logarithmic corrections due to these thermal fluctuations. We can then analyze the effect of such a corrected value of entropy on the stability of a charged dilatonic black Saturn.

\end{document}